# Strategies and Resources for teaching your physics course online on short notice

Chandralekha Singh

Department of Physics and Astronomy, University of Pittsburgh, Pittsburgh, PA 15260

**Abstract.** This article summarizes strategies and resources for teaching physics courses online on short notice.

Due to the COVID-19 pandemic, many of us teaching physics as well as those teaching other subjects in face-to-face brick-and-mortar classrooms have suddenly found themselves in an unprecedented situation: the rest of the term should immediately transition to a completely online format! Here I outline some strategies and resources that can help you and your colleagues in such a situation who have not had extensive time to think or plan for online instruction. It is important to recognize that everyone, including you and your students, is anxious and stressed by the circumstances, so being compassionate to yourself and your students is of paramount importance. I cannot emphasize enough that providing maximum flexibility to students is critical, particularly because they did not sign up for an online course and many of them may not have necessary resources, e.g., access to a computer with reliable internet connection or a quiet room at home or required time due to being sick or caring for a sick family member in order to complete all of the requirements of your online course. Everyone is trying to do their very best to salvage the situation, so whatever you and your students can accomplish is good. Communicating frequently and clearly with students is key.

There are many online resources, for example, Linda Strubbe and Sam McKagan's excellent crowd-sourced resources on PhysPort [1], that can be invaluable so you should definitely go over it in its entirety. Below, I summarize seven things to keep in mind while preparing for and executing your online physics courses, including labs:

1. **Keep the focus on the learning goals and objectives of your course.** For example, if your big picture goals are to help students learn to think like a physicist and help them develop problem solving, reasoning and meta-cognitive skills and become independent learners and excellent problem solvers, think carefully about strategies for how your online course will accomplish that [2-5]. Feel free to reduce the overall content coverage focusing on the central topics and effective approaches to engaging students and assessing their learning to meet your course goals in this online learning environment.
2. **For lecture-based courses, decide whether it is better to deliver your lectures synchronously or asynchronously.** Synchronous approach involves streaming your lectures live to students, e.g., via Zoom, BlueJeans, Skype or other platforms, and interacting with students during streaming similar in spirit to what you would do in a brick and mortar classroom. These types of platforms provide flexibility to record the streamed lecture that you can make available to students who could not join the live streaming. While the synchronous format gives students opportunity to ask questions and allows students to interact with each other and you, not requiring attendance during live stream and posting the recorded videos of live streamed sessions is essential to make sure students who do not have the resources or means to connect live are able to access relevant materials.
3. **Consider establishing virtual office hours** and have them at different times of the day so that students who are at home in different time zones can connect with you at a reasonable time. These live one-on-one or few-on-one sessions will give your students an opportunity to ask questions after they have had time to reflect on the material and work on homework. Moreover, it can be invaluable to have asynchronous discussion board on your learning management system, e.g., Blackboard or Canvas, where students and you can discuss what students are finding challenging and there is a record of those discussions for all students who may not be available at a certain time. Also, being available to students for office hour via virtual chat (instead of video) may be particularly helpful for students whose internet connections do not reliably support video interface.
4. **Consider using asynchronous pre-recorded lectures, created either by you or by others**. This way you can use all of the synchronous time with students for interactions, discussions and reflections. This approach is common in flipped mode [6] of teaching in which a large part of the meeting time with students is devoted to evidence-based active-engagement activities in the spirit of Just-in-time-teaching [7,8] and students interact with their peers and instructor after having gone over the pre-lectures and the corresponding pre-assessment tasks. In your online course, the actual meeting time will be virtual. Videoconferencing solutions such as Zoom have breakout rooms that allows you to assign students to different breakout rooms so that a smaller number of students in each breakout room can work with each other on the physics problems you assign using a virtual collaborative whiteboard for a certain length of time and then you can bring the students back into the same virtual room for a general discussion. In large classes, you may poll students by asking multiple-choice questions [9] that focus on your learning goals although it may be more difficult to engage students in peer interaction in this mode. Also, if you are pre-recording your own asynchronous lectures [10], make sure that you break your lecture into roughly 10-minute long sub-lectures and intersperse online assessments that you can grade students on, e.g., in the multiple-choice format, between each of these sub-lectures. This type of design is conducive to maintaining students' attention on each short sub-lecture and giving them an opportunity to assess their learning between different modules. Each of these pre-recorded sub-lectures can be e.g., voice over power point or similar to Khan Academy [11] (you will need a laptop or iPad with ability to write on it) and try to incorporate good visuals and if possible lecture demonstrations especially for introductory physics. If you are adopting asynchronous mode with pre-recorded video lectures, you can use the existing resources for introductory physics, e.g., FlipIt Physics [6], although it may cost money.
5. **For lab courses, take advantage of interactive virtual labs, simulations, and journal articles.** There are many such virtual labs (e.g., see [12-16], some of them are free while others may cost money beyond 30-day period). Articles in the *American Journal of Physics* (AJP) and *The Physics Teacher* (TPT) can be great resources in online teaching not only in lecture-based courses but particularly for your lab courses at all levels. For example, there are many experiments that have been discussed in a pedagogical manner in AJP and TPT. In these articles, e.g., instructors have often shared insights about classic experiments, e.g., single photon experiments for which video data are available as in this article [17], Millikan oil drop experiment [18], muon decay [19], and many others. You could ask students to read some of these articles about the experiments and then write about the aspects of those experimental set up that made them effective, how things evolved in that field and how trouble shooting was done, what the experimental errors were, their implications to physics in general and various other issues based upon your goals. You can also get together electronically with students and discuss what they got out of those papers and assess them on their writings and discussions. If possible, combine these tasks with students playing with related interactive simulations and analyzing data about these types of reflective assignments associated with journal articles. Similarly, in upper-level lecture classes, reading articles related to a topic that students would otherwise have learned via lecture would be an innovative approach to initiate a discussion via an electronic platform. For example, AJP and TPT articles often provide nice overview of the field including common student difficulties that can make it easier for students to understand the concepts in corresponding chapters in the textbook and issues associated with them. This pedagogical approach commensurate with your course goals can help students learn to read and reflect upon journal articles (good for becoming a lifelong learner) and enjoy the whole experience.

6. **Remind yourself that these are extraordinary circumstances and feel free to change assessment approach and be considerate.** It is ok to change assessment strategies as well as grading rubric and adjust the weight on the materials covered before and after going online. For example, it is ok for you to reduce the weight on the final exam or even eliminate the traditional final exam in favor of many low-stakes assessments, final projects and online presentations (that can be pre-recorded or can be synchronous so that students can field questions from their peers and you). You should consider giving students the opportunity to work in groups on these projects and online presentations since working with others can reduce isolation (otherwise, isolation can increase anxiety) and have students benefit from interactions. You can come up with novel group projects that meet the goals of your course in lieu of the final exam especially for your upper-level courses that require students to work in groups but have some individual accountability built into them (e.g., all students must present some part of their project individually and answer questions by peers and instructor). If you must give final exams that students will do at their own pace at home, use an honor code. Try to be extremely considerate to students since they may not have resources at home to take advantage of offering in the online learning environment. Be inclusive and think about whether it is appropriate and equitable to give students who cannot do the work due to constraints an incomplete so they can make up later or modify requirements for them commensurate with their constraints so that they can finish with everyone. Consider not giving a grade lower than what they would have gotten based upon their performance on the course thus far before going online

7. **Remember that technology is a tool and not the goal.** Make sure the focus is always on your students and their learning based upon your course goals and personalize learning as much as possible in this online environment so that students who are already disadvantaged in many ways are not penalized further. Share your ideas with your colleagues and help each other. We would have learned a lot about online learning at the end of this challenging period!

**References**

[1] https://www.physport.org/recommendations/Entry.cfm?ID=119906